\documentclass{Interspeech2024}

\interspeechcameraready

\usepackage{url}            
\title{Improving Robustness of LLM-based Speech Synthesis by Learning Monotonic Alignment}

\name{$^*$Paarth}{Neekhara}
\name{$^*$Shehzeen}{Hussain}
\name{Subhankar}{Ghosh}
\name{Jason}{Li}
\name{Rafael}{Valle}
\name{Rohan}{Badlani}
\name{Boris}{Ginsburg}

\address{NVIDIA Corporation, Santa Clara, CA, USA}
\email{\{pneekhara, shehzeenh, subhankarg, jasoli\}@nvidia.com}


\begin{document}

\maketitle

\begin{abstract}
Large Language Model (LLM) based text-to-speech (TTS) systems have demonstrated remarkable capabilities in handling large speech datasets and generating natural speech for new speakers.
However, LLM-based TTS models are not robust as the generated output can contain repeating words, missing words and mis-aligned speech (referred to as hallucinations or attention errors), especially when the text contains multiple occurrences of the same token. We examine these challenges in an encoder-decoder transformer model and find that certain cross-attention heads in such models implicitly learn the text and speech alignment when trained for predicting speech tokens for a given text. To make the alignment more robust, we propose techniques utilizing CTC loss and attention priors that encourage monotonic cross-attention over the text tokens. Our guided attention training technique does not introduce any new learnable parameters and significantly improves robustness of LLM-based TTS models.~\footnote{
Audio Examples: \url{https://t5tts.github.io/} \\
$^*$Denotes equal contribution
}

\end{abstract}

\vspace{-2mm}
\section{Introduction}

Large language models (LLMs) have revolutionized the landscape of deep generative AI with their unprecedented ability to generate coherent and contextually rich content across diverse domains.
In LLM-based generative models, data is quantized into discrete tokens, which allows the formulation of data synthesis as a language modeling task.
Transformer architectures such as GPT~\cite{radford2018improving} (decoder-only) and T5~\cite{raffel2020exploring} (encoder-decoder) are trained to autoregressively generate discrete tokens given a prompt, leading to a unified architecture that can be adapted across various data domains and synthesis tasks.
Particularly in the speech domain, there has been a recent surge in the use of LLMs for various speech synthesis applications such as text-to-speech (TTS) and speech-to-speech translation tasks~\cite{borsos2023audiolm,wang2023neural,wang2023speechx,yang2023uniaudio}.

TTS synthesis has been traditionally treated as a cascaded problem with intermediate mel-spectrogram representation that is typically modelled as a regression task~\cite{tacotron,lancucki2021fastpitch,neekhara2021expressive,mellotron,casanova2022yourtts}.
However, discrete neural audio codecs~\cite{encodec,dac_kumar2024high,zeghidour2021soundstream} have emerged as a promising intermediate audio representation, that not only preserve audio fidelity at a high compression rate, but are also suitable for training autoregressive transformer-based LLMs.
Audio LLMs~\cite{borsos2023audiolm,wang2023neural,wang2023speechx} have gained traction for their ability to generate audio seamlessly, eliminating the necessity for additional duration and pitch prediction models. Moreover, LLM-based speech synthesis models can scale up to large speech datasets 
and be prompted in diverse ways to perform tasks like zero-shot speech synthesis, multilingual speech synthesis and other audio generation tasks besides speech.
Despite their remarkable achievements, LLM-based TTS models suffer from attention errors resulting in mis-aligned speech, 
repeating and missing words, analogous to hallucinations~\cite{texthallucinationsurvey,azamfirei2023large} exhibited by LLMs in the text domain. This issue becomes more prominent when the input text is challenging and contains repeating words. For certain inputs, the probabilistic autoregressive inference of LLM-based TTS models can result in looping or infinite silences~\cite{song2024ella}. This issue makes LLM-based TTS models unreliable for real-world applications.

In our work, we investigate this hallucination issue and find that attention layers of LLM-based TTS models learn an implicit alignment between text and speech tokens when trained using the next-token prediction objective. 
In encoder-decoder transformers, the TTS alignment is learned in certain cross-attention heads of the decoder; while in decoder-only models, the alignment is learned in the self-attention layers.
Since the implicitly learned alignment in attention layers is unconstrained during training, it is not strictly monotonic which results in mis-aligned synthesis during inference. 
To address this challenge, we propose a learning procedure that encourages monotonic alignment in the attention layers of LLM-based TTS models, resulting in significantly more robust TTS models without modifying the architecture or introducing new parameters.

We design a TTS model based on 
an encoder-decoder T5~\cite{raffel2020exploring} transformer architecture, which takes text and audio tokens of a reference audio as input and autoregressively predicts the audio tokens of the target audio from the decoder. 
To improve robustness of the TTS model, we propose a technique to guide the cross-attention head of the T5 model using a static attention prior and alignment loss that encourages monotonic attention over the text input. Our experiments demonstrate that the proposed technique significantly improves intelligibility of the synthesized audio especially for challenging text inputs. The key contributions of our work are as follows:

\begin{itemize}
  \item We propose an encoder-decoder transformer model for TTS synthesis. To the best of our knowledge, this is the first attempt at synthesizing multi-codebook neural audio codecs with an encoder-decoder architecture.
  \item We develop an alignment learning technique to guide the cross-attention heads in our TTS model to learn monotonic alignment. Incorporating our proposed technique reduces Character Error Rate (CER) of synthesized speech from $9.03\%$ to $3.92\%$ on challenging texts.
  \item We compare audio codec models based on Residual Vector Quantization and Finite Scalar Quantization (FSQ). FSQ codecs not only improve audio quality but also simplify the data representation by allowing parallel codebook prediction.
  
\end{itemize}



\begin{figure*}[tp]
\vspace{-8mm}
    \centering
    \includegraphics[width=0.9\textwidth]{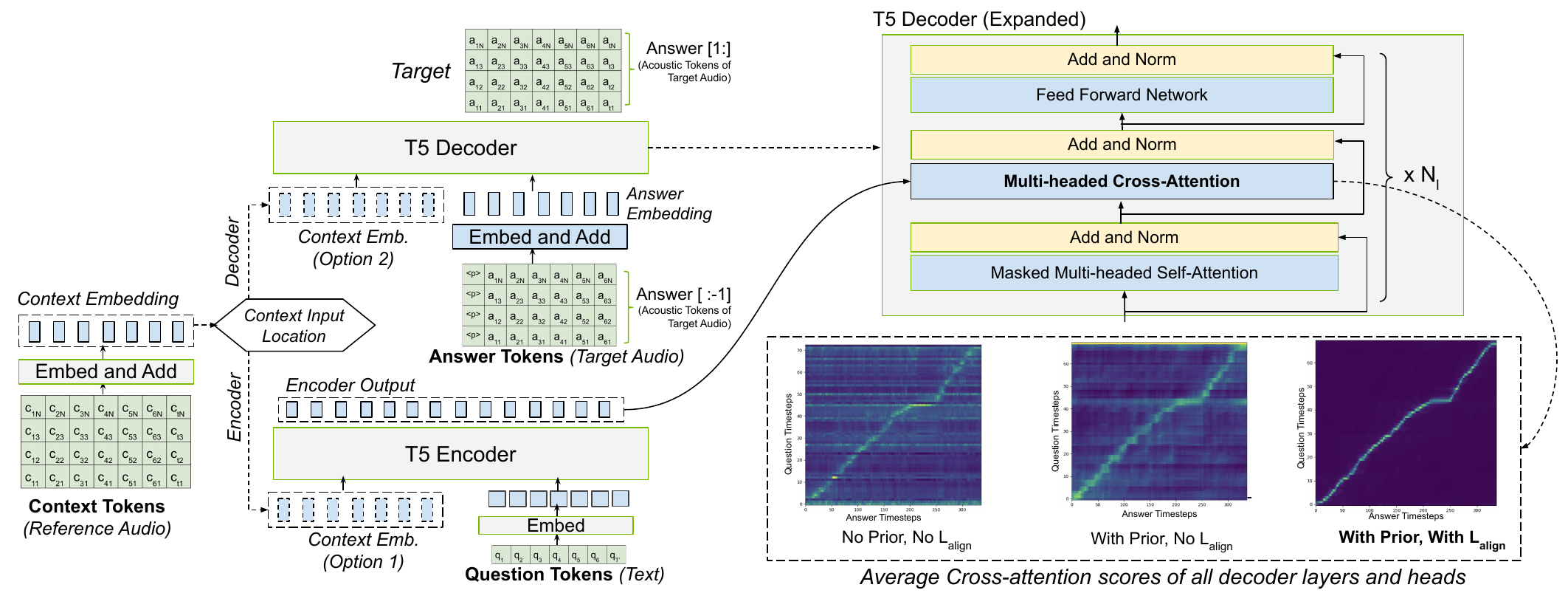}
\vspace{-2mm}    
    \caption{Model Overview: (Left) The T5-TTS model takes as input text tokens and acoustic codes of reference audio and predicts the acoustic codes of the target audio. The figure shows both context input location options. (Right) The cross-attention scores implicitly learn text and speech alignment, but can be guided to learn more robust alignment with attention prior and alignment loss $L_{\textit{align}}$}
\vspace{-4mm}
    \label{figs:model_overview}
\end{figure*}

\section{Related Work}



AudioLM~\cite{borsos2023audiolm} pioneered the task of training a decoder-only LLM on discretized audio tokens from a neural codec model, for high-quality speech synthesis. 
Following this, several solutions utilizing decoder-only transformer architectures have been proposed such as VALL-E,  UniAudio, Bark, SpeechX~\cite{wang2023neural,zhang2023speak,yang2023uniaudio,wang2023speechx}. They frame audio generation as an autoregressive language modeling task using multiple discrete codebooks. 
Alternatively, SpeechT5~\cite{ao2022speecht5} proposes an encoder-decoder architecture for sequence to sequence translation using a unified discrete representation of text and speech.
However, SpeechT5 similar to other synthesis models based on SSL representations~\cite{huang2022s3prl,hussain2023ace}, does not utilize multi-codebook audio representations. 

In the aforementioned transformer-based TTS models, the alignment between audio and phoneme sequences is entirely learned implicitly through the attention mechanisms in the transformer. This introduces potential instability in the form of hallucinations, since the alignment is not constrained to capture the monotonic dependencies of audio and text tokens~\cite{wang2023neural,song2024ella}. 
Prior research~\cite{tachibana2018efficiently,shih2021rad,onealign} on non-LLM spectrogram generation models have proposed solutions to learn stricter alignment between text and speech tokens by constraining the encoder-decoder attention layers in CNN-based TTS models and LSTM-based models such as Tacotron~\cite{tacotron} and Flowtron~\cite{valle2020flowtron}. While these techniques show promising results, they cannot be directly applied to transformer-based models which contain multiple cross-attention layers and multiple heads per layer, and generate discrete codes as opposed to continuous spectrograms.



\section{Methodology}

Our TTS model is an encoder-decoder LLM that is trained to predict acoustic codes of the target speech given the tokenized text input and acoustic codes of a reference audio from the target speaker. 
In this section, we first describe the tokenized representations used for text and speech. 
Next, we describe our model architecture and prompting setup for TTS. Finally, we propose a training procedure to learn robust text and speech alignment in the LLM.

\subsection{Tokenization} 
\label{sec:tokenization}
\noindent \textbf{Speech:} We utilize neural audio codec models to convert the raw speech signal into a tokenized representation. Given an audio signal $y=y_1 \dots y_t$, a neural audio codec model outputs $C_{T \times N}=\textit{CodecModel(y)}$. $C_{T \times N}$ is a two dimensional acoustic matrix containing $m$-bit discrete codes, where $T$ is the downsampled length and $N$ is the number of codebooks per timestep. We consider three acoustic codec models: Encodec~\cite{encodec}, Dac~\cite{dac_kumar2024high} and an unpublished Finite Scalar Quantization (FSQ)~\cite{mentzer2023finite} based spectral codec model~\cite{langman2024spectral}. Both Encodec and Dac use Residual Vector Quantization (RVQ)~\cite{zeghidour2021soundstream}. Due to RVQ's hierarchical architecture, we follow MusicGen~\cite{copet2024simple}'s delay pattern scheme for modelling the 
codebook depdendecies in RVQ.
In contrast, spectral codec~\cite{langman2024spectral} has $N$ independent codebooks. This allows us to predict the $N$ codebooks parallely at each timestep without using additional models or a delay pattern. To the best of our knowledge, we propose the first LLM that can parallely predict all $N$ codebooks and achieve high quality speech synthesis.


\noindent \textbf{Text:} For text, we use two tokenization schemes: sentence-piece~\cite{kudo2018sentencepiece} and phonemes. Sentence-piece tokens allows us to leverage pretrained text LLMs. To allow phoneme tokens as input, we expand the vocabulary and embedding space of the pretrained text-LLM to include additional tokens for phonemes. We train a single model to perform both phoneme to speech and sentence-piece to speech by prepending the text with the task prompt ``Phoneme TTS'' or  ``Text to Speech'' respectively. 
\label{codec}

\vspace{-2mm}
\subsection{Model Overview}

Our model is based on the T5 architecture~\cite{raffel2020exploring}, with additional embedding layers and prediction heads to adapt it for TTS task.
T5 is an encoder-decoder model, where the encoder is a non autoregressive bi-directional transformer and the decoder is an autoregressive transformer. 
Both the encoder and decoder networks contain $N_l$ transformer layers. 
Each layer within the encoder is composed of a self-attention module and a fully connected feed-forward network. In the decoder network, each layer adds an additional cross-attention module which performs multi-headed attention over the encoder's output.

To perform multi-speaker TTS, the model takes as input the tokenized text (\textit{question}), and the acoustic tokens of a reference audio from the target speaker (\textit{context}); and outputs the acoustic tokens of the target audio (\textit{answer}). 
We consider two design options in our experiments: feeding the \textit{context} audio tokens to the encoder network with the \textit{question}, or to the decoder network before the \textit{answer}. We discuss the trade-offs between these two designs in Section~\ref{sec:experiments}.

Note that the context and answer are represented by $N$ codebooks per timestep. 
To embed such tokens, we maintain $N$ embedding tables ($\texttt{EmbedA}_i$) of size $(2^m \times \textit{embDim})$ where $m$ is the bit-width of the acoustic codes ($m=10$ for all acoustic tokens in this work).  At each timestep, the acoustic embedding is derived by referencing and summing the embeddings from each of the $N$ codebooks.
Therefore, the acoustic embedding at a timestep $t$ is computed as:
$$e_t = \sum_{i=1}^{N}\texttt{EmbedA}_i(C[t,i]) $$
Similarly, at the decoder output, predictions for each of the $N$ codebooks are computed using a separate linear layer of size $h \times 2^m$, where $h$ is the hidden size of the transformer network. Therefore, for all timesteps and codebooks, we compute logits $y$ of size ${T \times N \times 2^m}$. Finally we calculate the cross entropy loss for next-token prediction as: $$L = \textit{CE}(\textit{SoftMax}(y), \textit{answer})$$
Note that unlike past work~\cite{wang2023neural,yang2023uniaudio}, our model does not use additional networks for handling multiple codebook predictions. Instead, we employ the delay pattern for representing RVQ tokens~\cite{copet2024simple} to model codebook dependencies.



\vspace{-1mm}
\subsection{Alignment Learning}
\label{sec:alignlearn}


When the T5 model is trained for TTS task using only the next token prediction loss, we observe that the 
attention-score matrix $A_{T\times M}$ in certain cross-attention heads, exhibits the learned text and speech alignment (where $T$ is the number of decoder timesteps and $M$ is the number of encoder timesteps).
That is, if we slice the attention-score matrix to include only the question time-steps, we observe higher attention-scores near the diagonal indicating the desirable monotonic alignment (Figure~\ref{figs:model_overview}).
However, attention errors in this implicitly learned alignment can cause missing or repeating words during inference, leading to hallucinations and inaccurate generations for challenging texts. 
Moreover, the alignment learning using only the next token prediction loss is often unstable and it can take several training iterations to learn a reasonable text and speech alignment, especially when training utterances are longer~\cite{song2024ella,onealign}. We extend prior work~\cite{tachibana2018efficiently,shih2021rad,onealign} and propose an alignment learning framework to guide multiple cross-attention heads of the T5 transformer model to learn robust alignment. 
\vspace{-2mm}
\subsubsection{Attention Prior}
\label{sec:attention}
\vspace{-1mm}
To accelerate alignment learning, 
during initial training we multiply the attention-score matrices in the cross-attention heads with a static 2D beta-binomial prior. 
The 2D beta-binomial prior is a near-diagonal heuristic matrix that is wider near the center and narrower near the corners. 
Multiplying the initially random attention matrices with such a prior, reduces the attention scores that are far-off the diagonal, providing a desirable monotonic initialization to the cross-attention scores. 

Consider the attention-score matrix between the decoder and encoder timesteps $A_{T\times M}^{\textit{l,h}}$,  of the $h^{\textit{th}}$ cross-attention head in decoder layer \textit{l}.
We generate a static 2d prior using the 2D beta-binomial distribution between the answer and question timesteps $P_{T' \times M'}$ where $T'$ is the number of time frames in the answer tokens and $M'$ is the number of question (text) timesteps. 
Given this prior, we obtain the re-scaled attention scores as: $${A}_{T\times M}^{\textit{l,h}}[a_s:a_e,q_s:q_e] \gets A_{T\times M}^{\textit{l,h}}[a_s:a_e,q_s: q_e] \odot  P_{T' \times M'}$$
where $q_s$ and $q_e$ indicate the start and end of the question timesteps ($M'=q_e - q_s$) and $a_s$ and $a_e$ indicate the start and end of the answer timesteps ($T'=a_e - a_s$). While $q_e=M, a_e=T$, the start timesteps $(q_s, a_s)$ for slicing depend on whether we pass context to the encoder or decoder. When passing context as input to the encoder, $q_s=\textit{context length}$, $a_s=0$ and when passing the context to the decoder, $q_s=0, a_s=\textit{context length}$.

We apply the prior to all cross-attention heads of each decoder layer.
Since we do not know the target audio length during inference  which is needed to compute the prior, we cannot use this prior during inference. Therefore, we apply the attention prior for the first $S_{1}$ training iterations. 
Then we linearly anneal the prior to an all ones matrix $J_{T' \times M'}$ from training step $S_1$ to $S_2$, and turn off the prior after step $S_2$. That is, for a training step $S$, where $S_1 \leq S \leq S_2$, the prior matrix is obtained as:
$$P_{T' \times M'}^S = ( (S_2 - S) \cdot P_{T' \times M'} + (S-S_1) \cdot J_{T' \times M'} ) /(S_2 - S_1)$$
This annealing procedure is necessary to ensure stability during training. 
Turning off the prior without annealing causes the loss curve to spike, since the decoder expects re-scaled attention scores for making valid predictions.
In our experiments, we set $S_1=8000$ and $S_2=15000$.
\vspace{-2mm}
\subsubsection{Alignment Loss}
\label{sec:ctc}
\vspace{-1mm}
The soft alignment matrix between the text and audio timesteps can be obtained by taking softmax of the sliced attention-score matrix over the text dimension:
$$A^{\textit{soft}_{l,h}}_{T'\times M'} = \textit{Softmax}( {A}_{T\times M}^{\textit{l,h}}[a_s:a_e,q_s:q_e] )$$ 
An $i^{\textit{th}}$ row in this matrix $A^{\textit{soft}_{l,h}}_{T'\times M'}[i,:]$ represents the attention probability distribution over all text timesteps for the given answer timestep $i$. 
If we sample a prediction from such a distribution at each answer timestep, it is desirable that
the resulting sequence of text timesteps is monotonic. 
Since the length of the answer is typically longer than the text,
there can be multiple valid monotnic reductions of the alignment matrix.

To encourage valid monotonic sampling from the alignment matrix, we calculate the likelihood of all possible monotonic reductions using the  Connectionist Temporal Classification (CTC) algorithm. That is, given the alignment matrix $A^{\textit{soft}_{l,h}}_{T'\times M'}$, we obtain the alignment loss for a decoder layer and head as: $$L_{\textit{align}}^{l,h} = \textit{CTCLoss}(A^{\textit{soft}_{l,h}}_{T'\times M'}, q_{M'})$$
where $q_{M'}=\{1, 2, \dots M'\}$ is the target monotonic sequence from $1$ to $M'$. 
We compute the total alignment loss for a set of cross-attention heads and layers $\mathbb{{P}}$ over which we wish to enforce monotonic alignment. That is,
$$ L_{\textit{align}} = \sum_{\substack{l,h \in \mathbb{{P}}}} L_{\textit{align}}^{l,h} $$

\noindent For set $\mathbb{{P}}$ we consider i) all cross-attention heads or ii) $2$ heads in each decoder layer. Observing no significant difference in intelligibility and robustness, for simplicity we apply $L_{\textit{align}}$ to all cross-attention heads in each layer in our experiments.





\section{Experiments}
\label{sec:experiments}

\subsection{Datasets and Models}
We train our T5-TTS models on a data-blend containing $1.8k$ hours of English TTS data from four datasets: the \emph{train-clean-360} subset of LibriTTS~\cite{zen2019libritts}, HiFiTTS~\cite{bakhturina21_interspeech}, a $1000$ hour subset of the LibriVox MLS dataset~\cite{pratap20_interspeech}, and a proprietary, 2-speaker, 63 hour dataset. 
The encoder and decoder transformer networks of our TTS model have $12$ layers and $12$ attention heads each, an embedding dimension of $768$, a feed-forward layer dimension of $4096$, and a dropout of $0.1$. This results in a total of $220$ million parameters excluding the embedding layers. We initialize our model weights with a pre-trained T5  checkpoint trained on Pile~\cite{pile}. 
To adapt the pre-trained text checkpoint for TTS, we make three modifications to the pretrained model: i) Expand the text vocabulary and corresponding embedding layers (initialized randomly) to include phoneme tokens. ii) Add randomly initialized embedding layers for each of the $N$ codebooks of the speech tokens. iii) Expand position embeddings to a maximum length of $1536$ which allows for an audio generation length of $20.5$ seconds using Encodec and $17.8$ seconds using Dac and spectral codec. 
We use a fixed context duration of $3$ seconds, where context is an alternate utterance from the speaker of the target utterance. 
We train each of our models with a batch size of $192$ distributed across $32$ NVIDIA A100 GPUs, for $250,000$ steps optimized with a fixed learning rate of $1e-4$ using AdamW optimizer. 
During inference, we use multinomial Top-k sampling with $k=80$ and temperature=$0.85$.

\vspace{-2mm}
\subsection{Results}
\label{sec:expal}
\vspace{-1mm}
\noindent \textbf{Alignment Learning:} 
To assess the efficacy of our alignment learning method (Section~\ref{sec:alignlearn}), 
we train three variants of our T5 TTS model using the spectral codec: 1) T5-TTS (No Prior, No $ L_{\textit{align}} ) $: trained without alignment learning method. 2) T5-TTS (W Prior, No $ L_{\textit{align}} ) $: trained with attention prior but not $L_\textit{align}$ and 3) T5 TTS (W Prior, W $ L_{\textit{align}} ) $: trained with attention prior and $L_{\textit{align}}$ applied to all cross-attention heads. 
In our initial experiments, the attention prior is crucial for training with $L_\textit{align}$. Without the prior and with $L_\textit{align}$, we obtain monotonic but unaligned attention maps.
leading to no speech synthesis. 

\setlength{\tabcolsep}{2pt}
\begin{table}[t]
\caption{\footnotesize{TTS results on seen and unseen speakers for different T5-TTS models. Lower CER(\%) \& WER(\%) indicate higher intelligibility. Higher SSIM indicates higher speaker similarity to ground-truth audio.}}
\vspace{-2mm}
\centering
\resizebox{\columnwidth}{!}{%
\begin{tabular}{c|ll|ccc}
\toprule
Eval Set & Model &  Context & CER $\downarrow$ & WER $\downarrow$ & SSIM $\uparrow$ \\
\midrule
& Ground Truth & & $1.03$& $2.08$  & $0.923$\\
LibriTTS & T5-TTS (No Prior, No $ L_{\textit{align}} ) $ & Encoder & $4.01$ & $6.74$ & $\mathbf{0.918}$ \\
(Seen & T5-TTS (W Prior, No $ L_{\textit{align}} ) $ & Encoder & $2.56$ & $4.68$ & $0.916$ \\
Speakers)  & T5-TTS (W Prior, W $ L_{\textit{align}} ) $ & Encoder & $2.16$ & $3.91$	& $0.911$ \\
 & T5-TTS (W Prior, W $ L_{\textit{align}} ) $ & Decoder & $\mathbf{1.69}$ & $\mathbf{3.60}$	& $0.900$ \\
\midrule
VCTK & Ground Truth & & $0.50$ & $0.83$ & $0.89$ \\
(Unseen & T5-TTS (W Prior, W $ L_{\textit{align}} ) $ & Encoder & $2.86$ & $4.66$ & $0.741$  \\
Speakers)& T5-TTS (W Prior, W $ L_{\textit{align}} ) $ & Decoder & $\mathbf{2.31}$ & $\mathbf{3.51}$ &	$\mathbf{0.779}$ \\
\bottomrule
\end{tabular} 
}
\vspace{-4mm}
\label{tab:ablation}
\end{table}

We evaluate the models on a set of seen and unseen speakers. For seen speakers, we use $200$ holdout utterances of the \textit{train-clean-360} set. For unseen speakers, we consider $200$ utterances from the VCTK~\cite{vctk} speakers: $20$ random speakers with $10$ utterances per speaker. 
For each utterance, we synthesize two audios using either the sentence piece text tokenizer or the phoneme tokenizer.
We evaluate the synthesized speech on intelligibility and speaker similarity. For intelligibility, we transcribe the synthesized audio through a Conformer-Transducer ASR model~\footnote{\fontsize{6}{6}\selectfont \url{https://hf.co/nvidia/stt_en_conformer_transducer_large}} and compute the CER and WER between the ASR transcript and the ground-truth text. 
For speaker similarity (SSIM), we compute the cosine similarity between the embeddings of the synthesized speech and target ground-truth audio obtained from WavLM~\cite{chen2022wavlm} speaker verification model~\footnote{\fontsize{6}{6}\selectfont \url{https://hf.co/microsoft/wavlm-base-plus-sv}}. 
We report the results in Table~\ref{tab:ablation}. 
While all three models achieve high speaker similarity for seen speakers, the intelligibility metrics improve as we incorporate attention prior and alignment loss during training. For unseen speakers, we observe a higher speaker similarity and intelligibility when the context is fed to the T5 decoder instead of the encoder.

\noindent \textbf{Challenging Texts and Comparison against Prior Work:} As shown in Table~\ref{tab:challenging}, the improvement in intelligibility is even more significant when we consider challenging text inputs with repeating words. 
We compare our models (using decoder context) with three open source LLM-based TTS models using the inference code and released checkpoints~\cite{ao2022speecht5,bark,vallexgithub}.
For this evaluation we consider a set of $100$ challenging texts and choose two seen speakers (male and female) from the voice presets of each model.
As observed, our best model outperforms the baseline models and prior LLM-based TTS models.
Additionally, we synthesize audio on $100$ texts from Harvard Sentences~\cite{harvardsentences} and conduct a Mean Opinion Score (MOS) evaluation on Amazon Mechanical Turk (Table~\ref{tab:challenging}). For MOS evaluation, each listener is presented with one audio sample and asked to rate the audio on a scale of $1$ to $5$ with $1$ point intervals. Each audio is rated by at least $10$ independent listeners. For $200$ audios per model, this results in a total of $2000$ evaluations per model. MOS with $95\%$ confidence intervals indicates our model outperforms prior LLM-based TTS models considered in our study. We encourage readers to listen to audio examples linked in the footnote of the first page.
\setlength{\tabcolsep}{3pt}
\begin{table}[t]
\caption{
\footnotesize{Comparison of different T5-TTS models against prior LLM-based TTS models. Intelligibility metrics (WER, CER, character insertions, deletions, substitutions (\%)) are evaluated on $100$ challenging texts. Naturalness MOS is calculated on subset of harvard sentences.}
}
\vspace{-2mm}
\centering
\resizebox{\columnwidth}{!}{
%
\begin{tabular}{l|rrrrr|c}
\multicolumn{1}{c}{} & \multicolumn{5}{c}{\emph{Intelligibility} $\downarrow$} & \multicolumn{1}{c}{\emph{Naturalness} $\uparrow$} \\
\toprule
Model & WER & CER & Ins & Del & Subs & MOS\\
\midrule
VALL-E-X~\cite{zhang2023speak} & $16.8$ &  $8.37$ & $1.88$ & $3.69$ & $2.80$ & $3.94 \pm 0.039$ \\
Bark~\cite{bark} & $19.1$ & $11.90$ & $1.33$ & $8.07$ &$2.50$ & $3.93 \pm 0.040$ \\
SpeechT5~\cite{ao2022speecht5} & $13.5$ & $6.14$ &	$\mathbf{0.94}$ & $3.92$ & $1.28$ & $3.98 \pm 0.042$\\
\midrule
T5-TTS (No Prior, No $ L_{\textit{align}} ) $ & $15.05$ & $9.03$ & $3.27$ & $4.19$ & $1.56$ & - \\
T5-TTS (W Prior, No $ L_{\textit{align}} ) $ & $13.04$ & $7.27$ & $3.47$ & $2.61$ & $1.18$ & - \\
\textbf{T5-TTS (W Prior, W }$ \mathbf{L_{\textit{\textbf{align}}}} \textbf{)}  $ & $\mathbf{9.22}$ & $\mathbf{3.92}$ & $1.24$ & $\mathbf{1.82}$ & $\mathbf{0.86}$ & $\mathbf{4.06 \pm 0.038}$  \\
\bottomrule
\end{tabular} 
}
\label{tab:challenging}
\end{table}



\noindent \textbf{Codec Choice:}
We train three T5-TTS models with alignment learning on the three audio codecs and report results on seen speakers in Table~\ref{tab:codeccomparison}.
We find that both spectral codec and Dac significantly outperform Encodec in terms of audio naturalness. 
Spectral codec streamlines training by independently predicting codebooks in parallel, unlike the delay pattern scheme needed for Encodec/Dac. Additionally, spectral codec enhances synthesized speech intelligibility, demonstrated by reduced CER/WER.

\setlength{\tabcolsep}{3pt}
\begin{table}[tp]
\vspace{-2mm}
\caption{
\footnotesize{Comparison of T5-TTS (W Prior, W $ L_{\textit{align}})$ models trained with different audio codecs. N: number of codebooks, FPS: Frames Per Second, TPS: Tokens Per Second. All codecs use 10-bit tokens.}
}
\vspace{-2mm}
\centering
\resizebox{0.9\columnwidth}{!}{%
\begin{tabular}{lrrr|cccc}
\toprule
Codec & FPS & N & TPS & CER $\downarrow$  & WER $\downarrow$  & SSIM $\uparrow$ & MOS $\uparrow$\\
\midrule
Encodec & 75 & 8 & 600 & $4.01$ & $7.20$ & $\mathbf{0.920}$ & $3.57 \pm 0.06$\\
Dac & 86 &  9 & 774 & $6.72$ & $9.65$ & $0.910$ & $\mathbf{ 3.92 \pm 0.05 }$ \\
Spectral codec & 86 & 8 & 688 & $\mathbf{2.16}$ & $\mathbf{3.91}$	& $0.911$ & $\mathbf{ 3.91 \pm 0.05 }$ \\
\bottomrule
\end{tabular} 
}
\vspace{-6mm}
\label{tab:codeccomparison}
\end{table}

\section{Conclusion}
We present a T5-TTS model that can learn robust text and speech alignment without modifying the model architecture or requiring ground-truth text duration. We identify that attention heads in LLM-based TTS models implicitly learn text and speech alignment and can be guided to monotonically attend over the text input. 
Our experiments demonstrate that our alignment learning procedure improves the reliability of TTS synthesis, especially for challenging text inputs and outperforms prior LLM-based TTS models on both intelligibility and naturalness.

\section{Acknowledgements}
We would also like to thank Ryan Langman for developing the spectral codec model that was used in our TTS model.


\bibliographystyle{IEEEtran}
\bibliography{mybib}

\begin{thebibliography}{10}
\providecommand{\url}[1]{#1}
\csname url@samestyle\endcsname
\providecommand{\newblock}{\relax}
\providecommand{\bibinfo}[2]{#2}
\providecommand{\BIBentrySTDinterwordspacing}{\spaceskip=0pt\relax}
\providecommand{\BIBentryALTinterwordstretchfactor}{4}
\providecommand{\BIBentryALTinterwordspacing}{\spaceskip=\fontdimen2\font plus
\BIBentryALTinterwordstretchfactor\fontdimen3\font minus \fontdimen4\font\relax}
\providecommand{\BIBforeignlanguage}[2]{{%
\expandafter\ifx\csname l@#1\endcsname\relax
\typeout{** WARNING: IEEEtran.bst: No hyphenation pattern has been}%
\typeout{** loaded for the language `#1'. Using the pattern for}%
\typeout{** the default language instead.}%
\else
\language=\csname l@#1\endcsname
\fi
#2}}
\providecommand{\BIBdecl}{\relax}
\BIBdecl

\bibitem{radford2018improving}
A.~Radford, K.~Narasimhan, T.~Salimans, I.~Sutskever \emph{et~al.}, ``Improving language understanding by generative pre-training,'' \emph{OpenAI blog}, 2018.

\bibitem{raffel2020exploring}
C.~Raffel, N.~Shazeer, A.~Roberts, K.~Lee, S.~Narang, M.~Matena, Y.~Zhou, W.~Li, and P.~J. Liu, ``Exploring the limits of transfer learning with a unified text-to-text transformer,'' \emph{The Journal of Machine Learning Research}, 2020.

\bibitem{borsos2023audiolm}
Z.~Borsos, R.~Marinier, D.~Vincent, E.~Kharitonov, O.~Pietquin, M.~Sharifi, D.~Roblek, O.~Teboul, D.~Grangier, M.~Tagliasacchi \emph{et~al.}, ``Audiolm: a language modeling approach to audio generation,'' \emph{IEEE/ACM Transactions on Audio, Speech, and Language Processing}, 2023.

\bibitem{wang2023neural}
C.~Wang, S.~Chen, Y.~Wu, Z.~Zhang, L.~Zhou, S.~Liu, Z.~Chen, Y.~Liu, H.~Wang, J.~Li \emph{et~al.}, ``Neural codec language models are zero-shot text to speech synthesizers,'' \emph{arXiv:2301.02111}, 2023.

\bibitem{wang2023speechx}
X.~Wang, M.~Thakker, Z.~Chen, N.~Kanda, S.~E. Eskimez, S.~Chen, M.~Tang, S.~Liu, J.~Li, and T.~Yoshioka, ``Speechx: Neural codec language model as a versatile speech transformer,'' \emph{arXiv:2308.06873}, 2023.

\bibitem{yang2023uniaudio}
D.~Yang, J.~Tian, X.~Tan, R.~Huang, S.~Liu, X.~Chang, J.~Shi, S.~Zhao, J.~Bian, X.~Wu \emph{et~al.}, ``Uniaudio: An audio foundation model toward universal audio generation,'' \emph{arXiv:2310.00704}, 2023.

\bibitem{tacotron}
Y.~Wang, R.~Skerry-Ryan, D.~Stanton, Y.~Wu, R.~J. Weiss, N.~Jaitly, Z.~Yang, Y.~Xiao, Z.~Chen, S.~Bengio \emph{et~al.}, ``Tacotron: Towards end-to-end speech synthesis,'' in \emph{INTERSPEECH}, 2017.

\bibitem{lancucki2021fastpitch}
A.~{\L}a{\'n}cucki, ``Fastpitch: Parallel text-to-speech with pitch prediction,'' in \emph{ICASSP}, 2021.

\bibitem{neekhara2021expressive}
P.~Neekhara, S.~Hussain, S.~Dubnov, F.~Koushanfar, and J.~McAuley, ``Expressive neural voice cloning,'' in \emph{Asian Conference on Machine Learning}.\hskip 1em plus 0.5em minus 0.4em\relax PMLR, 2021.

\bibitem{mellotron}
R.~Valle, J.~Li, R.~Prenger, and B.~Catanzaro, ``Mellotron: Multispeaker expressive voice synthesis by conditioning on rhythm, pitch and global style tokens,'' \emph{ICASSP}, 2020.

\bibitem{casanova2022yourtts}
E.~Casanova, J.~Weber, C.~Shulby, A.~Junior, E.~G{\"o}lge, and M.~A. Ponti, ``Yourtts: Towards zero-shot multi-speaker tts and zero-shot voice conversion for everyone,'' in \emph{ICML}.\hskip 1em plus 0.5em minus 0.4em\relax PMLR, 2022.

\bibitem{encodec}
A.~D{\'e}fossez, J.~Copet, G.~Synnaeve, and Y.~Adi, ``High fidelity neural audio compression,'' \emph{Transactions on Machine Learning Research}, 2023.

\bibitem{dac_kumar2024high}
R.~Kumar, P.~Seetharaman, A.~Luebs, I.~Kumar, and K.~Kumar, ``High-fidelity audio compression with improved rvqgan,'' \emph{Advances in Neural Information Processing Systems}, 2024.

\bibitem{zeghidour2021soundstream}
N.~Zeghidour, A.~Luebs, A.~Omran, J.~Skoglund, and M.~Tagliasacchi, ``Soundstream: An end-to-end neural audio codec,'' \emph{IEEE/ACM Transactions on Audio, Speech, and Language Processing}, 2021.

\bibitem{texthallucinationsurvey}
Z.~Ji, N.~Lee, R.~Frieske, T.~Yu, D.~Su, Y.~Xu, E.~Ishii, Y.~J. Bang, A.~Madotto, and P.~Fung, ``Survey of hallucination in natural language generation,'' \emph{Association for Computing Machinery}, 2023.

\bibitem{azamfirei2023large}
R.~Azamfirei, S.~R. Kudchadkar, and J.~Fackler, ``Large language models and the perils of their hallucinations,'' \emph{Critical Care}, 2023.

\bibitem{song2024ella}
Y.~Song, Z.~Chen, X.~Wang, Z.~Ma, and X.~Chen, ``Ella-v: Stable neural codec language modeling with alignment-guided sequence reordering,'' \emph{arXiv:2401.07333}, 2024.

\bibitem{zhang2023speak}
Z.~Zhang, L.~Zhou, C.~Wang, S.~Chen, Y.~Wu, S.~Liu, Z.~Chen, Y.~Liu, H.~Wang, J.~Li \emph{et~al.}, ``Speak foreign languages with your own voice: Cross-lingual neural codec language modeling,'' \emph{arXiv:2303.03926}, 2023.

\bibitem{ao2022speecht5}
J.~Ao, R.~Wang, L.~Zhou, C.~Wang, S.~Ren, Y.~Wu, S.~Liu, T.~Ko, Q.~Li, Y.~Zhang \emph{et~al.}, ``Speecht5: Unified-modal encoder-decoder pre-training for spoken language processing,'' in \emph{Proceedings of the 60th Annual Meeting of the Association for Computational Linguistics (Volume 1: Long Papers)}, 2022, pp. 5723--5738.

\bibitem{huang2022s3prl}
W.~C. Huang, S.~W. Yang, T.~Hayashi, H.~Y. Lee, S.~Watanabe, and T.~Toda, ``S3prl-vc: Open-source voice conversion framework with self-supervised speech representations,'' in \emph{ICASSP}, 2022.

\bibitem{hussain2023ace}
S.~Hussain, P.~Neekhara, J.~Huang, J.~Li, and B.~Ginsburg, ``Ace-vc: Adaptive and controllable voice conversion using explicitly disentangled self-supervised speech representations,'' in \emph{ICASSP}, 2023.

\bibitem{tachibana2018efficiently}
H.~Tachibana, K.~Uenoyama, and S.~Aihara, ``Efficiently trainable text-to-speech system based on deep convolutional networks with guided attention,'' in \emph{ICASSP}, 2018.

\bibitem{shih2021rad}
K.~J. Shih, R.~Valle, R.~Badlani, A.~Lancucki, W.~Ping, and B.~Catanzaro, ``Rad-tts: Parallel flow-based tts with robust alignment learning and diverse synthesis,'' in \emph{ICML Workshop on Invertible Neural Networks, Normalizing Flows, and Explicit Likelihood Models}, 2021.

\bibitem{onealign}
R.~Badlani, A.~Łańcucki, K.~J. Shih, R.~Valle, W.~Ping, and B.~Catanzaro, ``One tts alignment to rule them all,'' in \emph{ICASSP}, 2022.

\bibitem{valle2020flowtron}
R.~Valle, K.~J. Shih, R.~Prenger, and B.~Catanzaro, ``Flowtron: an autoregressive flow-based generative network for text-to-speech synthesis,'' in \emph{ICLR}, 2020.

\bibitem{mentzer2023finite}
F.~Mentzer, D.~Minnen, E.~Agustsson, and M.~Tschannen, ``Finite scalar quantization: Vq-vae made simple,'' in \emph{ICLR}, 2023.

\bibitem{langman2024spectral}
R.~Langman, A.~Jukić, K.~Dhawan, N.~R. Koluguri, and B.~Ginsburg, ``Spectral codecs: Spectrogram-based audio codecs for high quality speech synthesis,'' 2024.

\bibitem{copet2024simple}
J.~Copet, F.~Kreuk, I.~Gat, T.~Remez, D.~Kant, G.~Synnaeve, Y.~Adi, and A.~D{\'e}fossez, ``Simple and controllable music generation,'' \emph{Advances in Neural Information Processing Systems}, vol.~36, 2024.

\bibitem{kudo2018sentencepiece}
T.~Kudo and J.~Richardson, ``Sentencepiece: A simple and language independent subword tokenizer and detokenizer for neural text processing,'' in \emph{Proceedings of the 2018 Conference on Empirical Methods in Natural Language Processing: System Demonstrations}, 2018.

\bibitem{zen2019libritts}
H.~Zen, V.~Dang, R.~Clark, Y.~Zhang, R.~J. Weiss, Y.~Jia, Z.~Chen, and Y.~Wu, ``Libritts: A corpus derived from librispeech for text-to-speech,'' \emph{INTERSPEECH}, 2019.

\bibitem{bakhturina21_interspeech}
E.~Bakhturina, V.~Lavrukhin, B.~Ginsburg, and Y.~Zhang, ``{Hi-Fi Multi-Speaker English TTS Dataset},'' in \emph{INTERSPEECH}, 2021.

\bibitem{pratap20_interspeech}
V.~Pratap, Q.~Xu, A.~Sriram, G.~Synnaeve, and R.~Collobert, ``{MLS: A Large-Scale Multilingual Dataset for Speech Research},'' in \emph{INTERSPEECH}, 2020.

\bibitem{pile}
L.~Gao, S.~Biderman, S.~Black, L.~Golding, T.~Hoppe, C.~Foster, J.~Phang, H.~He, A.~Thite, N.~Nabeshima, S.~Presser, and C.~Leahy, ``The {P}ile: An 800gb dataset of diverse text for language modeling,'' \emph{arXiv:2101.00027}, 2020.

\bibitem{vctk}
C.~Veaux, J.~Yamagishi, and K.~Macdonald, ``{CSTR VCTK} corpus: English multi-speaker corpus for {CSTR} voice cloning toolkit,'' 2017.

\bibitem{chen2022wavlm}
S.~Chen, C.~Wang, Z.~Chen, Y.~Wu, S.~Liu, Z.~Chen, J.~Li, N.~Kanda, T.~Yoshioka, X.~Xiao \emph{et~al.}, ``Wavlm: Large-scale self-supervised pre-training for full stack speech processing,'' \emph{IEEE Journal of Selected Topics in Signal Processing}, 2022.

\bibitem{bark}
\BIBentryALTinterwordspacing
SunoAI, ``Bark audio generation model,'' 2023. [Online]. Available: \url{https://github.com/suno-ai/bark}
\BIBentrySTDinterwordspacing

\bibitem{vallexgithub}
\BIBentryALTinterwordspacing
Songting, ``{VALL-E-X},'' 2023. [Online]. Available: \url{https://github.com/Plachtaa/VALL-E-X}
\BIBentrySTDinterwordspacing

\bibitem{harvardsentences}
``{IEEE} recommended practice for speech quality measurements,'' \emph{IEEE Transactions on Audio and Electroacoustics}, vol.~17, no.~3, pp. 225--246, 1969.

\end{thebibliography}

\end{document}